\documentclass[12pt,preprint]{aastex}

\shorttitle{Discovery of inward motion}
\shortauthors{Zhang et al.}

\begin{document}

\title{Discovery of inward moving magnetic enhancements in sunspot penumbrae}

\author{Jun Zhang\altaffilmark{1, 2}, S. K. Solanki\altaffilmark{2},
and J. Woch \altaffilmark{2}}

\altaffiltext{1}{National Astronomical Observatories, Chinese
Academy of Sciences, Beijing 100012, China;
zjun@ourstar.bao.ac.cn}
 \altaffiltext{2}{Max-Planck-Institut
f\"{u}r Sonnensystemforschung,
Max-Planck-Str. 2, 37191 Katlenburg-Lindau, Germany\\
email: solanki@mps.mpg.de, woch@mps.mpg.de}

\begin{abstract}

Sunspot penumbrae show a fine structure in continuum intensity that
displays considerable dynamics. The magnetic field, in contrast,
although also highly structured, has appeared to be relatively
static. Here we report the discovery of inward moving magnetic
enhancements in the penumbrae of two regular sunspots based on time
series of SOHO/MDI magnetograms. Local enhancements of the LOS
component of the magnetic field in the inner part of the penumbral
region move inward to the umbra-penumbra boundary with a radial
speed of about 0.3 km s$^{-1}$. These local inward-moving
enhancements of the LOS component of the magnetic fields appear to
be relatively common. They are associated with dark structures and
tend to display downflows relatively to the penumbral background.
Possible explanations are discussed.
\end{abstract}


\section{INTRODUCTION}

Sunspot penumbrae are both, structured in a complex manner and
highly dynamic. They display fine structure in the form of dark and
bright fibrils \citep[the latter with dark cores;][]{sch02} and
bright points, called penumbral grains. Dynamic features associated
with the penumbral photosphere are the horizontally outward directed
Evershed flow, the outward motion of dark clouds that seem to
dominate the outer penumbra \citep{shi94} and the steady inward
motion of penumbral grains \citep[e.g.][]{mul76, sob99}, as well as
oscillations \citep[e.g.][]{mus76, blo07}. See \citet{sol03} for a
review.

The magnetic field, just like the brightness, also displays a
complex structure in the penumbra, with interlaced regions of
horizontal and more inclined field lying nearly parallel to
penumbral fibrils \citep{deg91, tit93}. These have been interpreted
in terms of horizontal flux tubes embedded in an inclined field
\citep[uncombed field;][]{sol93, bel04, bor05, bor06} and of
field-free gaps between the field lines \citep{sch06}.

Unlike the brightness structure, there has so far been little
evidence for significant, persistent changes of the fine-scale
magnetic structure of the penumbra. In fact, the magnetic pattern
has been found to change little over a period of an hour
\citep{so03}. The main exception is the outward motion of
magnetic enhancements that later move into the moat and become
Moving Magnetic Features (MMFs) \citep{sai05}. Here we present the
first observation of inward moving magnetic enhancements in
penumbrae and determine their continuum brightness signature.

\section{OBSERVATIONS}

The data sets analyzed here consist of magnetograms, Dopplergrams
and continuum images, obtained by the Michelson Doppler Imager
\citep[MDI,][]{sch95} onboard the {\it SOHO} spacecraft, operated in
its high-resolution observing mode with a spatial and temporal
sampling of 0.6$''$ and 1 minute, respectively. The targets are the
active regions, NOAA AR 0330 ($\mu$ = 0.984, which corresponds to a
heliocentric angle $\theta$ of 10$^{\circ}$) and NOAA AR 9697 ($\mu$
= 0.974, $\theta$ = 13$^{\circ}$). NOAA 0330 was observed between
2003 April 9, 14:05 UT and 2003 April 10, 00:49 UT, while NOAA 9697
was observed between 2001 Nov. 17, 18:00 UT and Nov. 18, 06:10 UT.
By treating the data as \citet{sai05} have done, i.e. selecting a
square region of 180$''$${\times}$180$''$, compensating solar
rotation and correcting for border effects, we follow the transit of
a sunspot through the MDI high-resolution area. All magnetograms,
Dopplergrams and continuum images are further co-aligned by
searching for the maximum of their correlation with respect to a
single reference magnetogram. This ensures that any proper motion of
the sunspot as a whole is removed, so that the center of gravity of
the spot remains roughly at the same position. Fig. 1 shows
continuum images of the two studied sunspots. The rectangular frame
outlines a subfield (see Fig. 2), and the solid lines `AB' and `CD'
cross the penumbra from the inner to outer boundary at locations at
which we present time slices. In the following we discuss the
dynamics of the field, the brightness and the velocity in these
locations.

\section{RESULTS}

In Fig. 2 we display a time sequence of the continuum images (left
column) and the magnetograms (right column) in the subfield marked
in the upper panel of Fig. 1. The light yellow continuum intensity
contours outline a dark feature, the white ones the umbral boundary
(see the arrow in the continuum image at 17:34 UT). These contours
are overplotted onto the corresponding magnetograms. The brightness
of the dark feature is about 80\% of the average brightness in the
penumbra and it corresponds to an enhancement in the magnetogram
signal, $B_{||}$ (i.e. the LOS component of the magnetic field) seen
in the right panels of Fig. 2 as a finger of enhanced brightness.
The $B_{||}$ enhancement and the associated dark feature first
appear in the inner half of the penumbra and move to the inner
penumbral boundary with an average speed of 0.3 km s$^{-1}$ from
17:34 UT to 19:36 UT. The arrow in the continuum image at 18:35 UT
denotes the motion direction of the feature. Finally this feature
intrudes into the umbra at 20:36 UT. The fact that this feature
appears isolated in brightness, but only as an intrusion of high
field into the penumbra has to do with the strong radial gradient in
the magnetogram signal. When this gradient is removed (Fig. 3) the
magnetogram enhancement becomes clearly visible.

Fig. 3 shows time-slice maps taken from the continuum images (left),
the magnetograms (middle), and the Dopplergrams (right),
respectively. The X-axis is the distance ($\sim$14 Mm) from `A' to
`B' (see Fig. 1), which cuts the penumbra from the inner to the
outer boundary, and the Y-axis is the time from 2003 April 9, 14:05
UT to April 10, 00:45 UT. In order to better reveal small-scale
(moving) features, we make a 2-D quadratic polynomial fit to each of
the upper frames. After the removal of this polynomial fit (lower
frames) considerable dynamic structure becomes visible, with
different features in the inner (left of the vertical white lines)
and the outer penumbra. In the outer penumbra, a large magnetic
feature is seen to move steadily outward with a speed of 0.25 km
s$^{-1}$ for about 7 hr. Upon leaving the sunspot it becomes an MMF.
The dotted line in the lower magnetogram time-slice shows its
trajectory. In the outer penumbra there is also a possible hint of
small magnetic features that move outward with an average speed
larger than 1.2 km s$^{-1}$ (the arrows in the lower middle panel
point in the directions of motion). These features merge with the
large one and continue their outward motion together.

In the inner penumbra, however, magnetic features with larger
$B_{||}$, denoted `M1' and `M2', move inward towards the umbra.
The feature `M1', already present at the start of the MDI
observation period on 2003 April 9, 14:05 UT, persisted for 5 hr
while moving to the inner boundary with a mean speed of 0.3 km
s$^{-1}$. About two hours later, the second feature `M2' appeared.
It moved to the inner boundary with a speed of 0.35 km s$^{-1}$, and
remained visible for 2-2.5 hr.

These inward moving features are slightly darker than average (they
avoid bright features), and the Dopplergram counterparts have a
tendency to show a relative red shift (brighter regions in the
right-most panels of Fig. 3). In the inner penumbral region (see the
lower frames of Fig. 3), there is a relationship between continuum
intensity and magnetic flux density residuals (with a correlation
coefficient of $-$0.52), as well as between continuum intensity and
Doppler shift residuals \citep[cf.,][]{sch99}, with a correlation
coefficient of $-$0.55.

Fig. 4 shows time-slice maps along the line CD (active region NOAA AR
9697). In the outer penumbra a feature that later becomes an MMF (see
the shorter arrow in the middle panel) moves outward to the moat
around the sunspot. In the inner
penumbra, a larger magnetic feature denoted `M' moves to the inner
boundary with an average speed of 0.3 km s$^{-1}$. The longer arrow
indicates the direction of motion. This feature is obviously darker
than average and is associated with a slight average redshift relative
to its surroundings.

Inward motion of features displaying an enhanced magnetogram
signal is also found at other locations in the studied penumbrae,
as well as in the main sunspot of NOAA AR 8375 in November 1998,
which we also analyzed and which confirms the results found for the
two sunspots presented here.

\section{DISCUSSION AND CONCLUSIONS}

We report on the first detection of inward moving enhancements of
the magnetogram signal in the inner halves of sunspot penumbrae,
associated with a local darkening and possibly a weak downflow.
These features can be followed right to the umbral boundary. This
phenomenon appears to be quite common, since we noticed it at
numerous locations in the penumbrae of three different mature
regular sunspots. We note that an enhancement in the magnetogram
signal can be produced by an enhanced field strength, by a field
aligned more strongly with the LOS (which, for a sunspot close to
disc center, is equivalent to a more vertical field), and
conceivably also by spatially unresolved variations of continuum
brightness or LOS velocity. The MDI data do not allow to distinguish
easily between these possibilities.

\citet{lan05} pointed out that
dark penumbral cores are associated with the more horizontal
component of the magnetic field, while the bright component of
filaments is associated with the more vertical component of the
magnetic field. The magnetogram signal is lower in the dark cores
than in the bright parts of penumbral filaments. This suggests that
they described other features than we have studied.
An enhanced $B_{||}$ associated with a darkening in the penumbra is
more likely to be related to the larger scale ``spines'' of more
vertical field found by \citet[][cf. Bello Gonzalez et al.
2005]{lit93}. It is interesting that Bello Gonzalez et al. (2005)
note the presence of downflows in the spines, which strengthens the
correspondence with the inward moving features that we see.

\citet{s1998, sch98} have proposed that the bright features in
penumbrae moving toward the umbra in the inner penumbra are the
locations where hot flux tubes emerge that become horizontal further
out in the penumbra. It is not clear in this model whether the
inward moving features are associated with an enhancement or
depression of $B_{||}$. More promising is the more recent version of
this model by \citet{sch03}, who finds that the horizontal flux tube
develops into a sea serpent, with the innermost visible part of it
moving towards the umbra, while in the outer penumbra the magnetic
enhancements would move outwards and presumably become visible later
as MMFs. The first footpoint, at which the flux tube and the
Evershed flow it carried dives below the surface again, should show
a downflow and may well be cool. This model would thus explain the
remarkable divergence between the inner and outer parts of the
penumbra, with magnetogram enhancements in the outer penumbra moving
outwards, those in the inner penumbra inwards.

We are keenly aware of the limitations imposed by the comparatively
low spatial resolution of MDI, even in its high resolution mode.
Future observations at higher spatial resolution are likely to
uncover more of the nature of these features. An analysis of such
observations is planned.

\acknowledgements

The authors are indebted to the {\it SOHO}/MDI team for providing
the data. {\it SOHO} is a project of international cooperation
between ESA and NASA. This work is supported by the National Natural
Science Foundations of China (G10573025 and 40674081), the CAS
Project KJCX2-YW-T04, and the National Basic Research Program of
China under grant G2006CB806303.

\clearpage

\begin{figure}
\epsscale{0.50}
\caption{Continuum intensity images
from {\it SOHO}/MDI showing the main sunspots belonging to active
regions NOAA AR 0330 (top) and AR 9697 (bottom). The field-of-view
of both frames is about 70$''$${\times}$70$''$. The white window
denotes a sub-area of the inner penumbral region where a dark
feature with an enhanced magnetogram signal moves towards the umbra
(see Fig. 2). The solid lines `AB' and `CD' cut the penumbrae from
the inner to outer boundary. Time slices of measured quantities
along these lines are shown in Figs. 3 and 4. The arrows point to
disk center. \label{fig1}}
\end{figure}

\clearpage

\begin{figure}
\epsscale{0.70}
\caption{Time sequence of continuum
images (left column) and the corresponding magnetograms (right
column) in the field-of-view (9$''$${\times}$6$''$) of the window
marked in Fig. 1. The continuum intensity contours are overplotted
also on the corresponding magnetograms. The dynamic ranges are from
1400 counts pixel$^{-1}$ to 2200 counts pixel$^{-1}$ for the
continuum images and from 400 G to 1100 G for the magnetograms. The
arrows are described in the text. \label{fig2}}
\end{figure}

\clearpage

\begin{figure}
\epsscale{0.50}
\caption{Time-slice maps taken from
the continuum images (left), from the corresponding magnetograms
(middle) and Dopplergrams (right), respectively. For each map, the
X-axis represents the distance ($\sim$14 Mm) from `A' to `B' (see
Fig. 1), which cuts the penumbra from the inner to outer boundary
(left to right). The Y-axis denotes time, running from 2003 April 9,
14:05 UT to April 10, 00:45 UT (from bottom to top). The upper
frames show original data, with the colour scale ranging from 1200
counts pixel$^{-1}$ to 2400 counts pixel$^{-1}$ for the continuum
map, from 150 G to 1000 G for the magnetogram map, and from $-$180
m~s$^{-1}$ (black, blue shift) to 180 m~s$^{-1}$ (white, red shift)
for the Dopplergram map. The lower frames display relative signals
which are obtained after subtracting a second-degree polynomial
surface fit from the original data. The dynamic ranges are from
$-$100 counts pixel$^{-1}$ to 100 counts pixel$^{-1}$ for the
continuum map, from $-$100 G to 100 G for the magnetogram map, and
from $-$120 m~s$^{-1}$ to 120 m~s$^{-1}$ for the Dopplergram map.
The 80 G contour overplotted on all time slices, refers to the
filtered magnetogram signal in the lower-middle map. The vertical
white lines on the lower maps separate inward moving magnetic
features (denoted by `M1' and `M2') from those moving outward. The
dotted line and the arrows are described in the text. \label{fig3}}
\end{figure}

\begin{figure}
\epsscale{0.70}
\caption{Same as the lower 3 frames
in Fig. 3 but for the main spot of active region NOAA AR 9697. The
X-axis represents the distance ($\sim$14 Mm, from `C' to `D' in Fig.
1) from the inner to outer penumbral boundary (left to right). The
Y-axis denotes the time running from 2001 Nov. 17, 19:51 UT to Nov.
18, 06:00 UT (bottom to top). `M' denotes an inward moving magnetic
feature, the black arrow its direction of motion, the white arrow an
outward moving feature. \label{fig4}}
\end{figure}

\clearpage

\end{document}